\documentclass{vgtc}                          % final (conference style)
\graphicspath{{figures/}{pictures/}{images/}{./}} % where to search for the images

\usepackage{times}                     % we use Times as the main font
\usepackage{tabu}                      % only used for the table example
\usepackage{booktabs}                  % only used for the table example
\usepackage{lipsum}                    % used to generate placeholder text
\usepackage{mwe}                       % used to generate placeholder figures
\usepackage{balance}
\usepackage{wrapfig}
\usepackage{subfigure}
\usepackage{subcaption}
\usepackage{mathptmx}                  % use matching math font
\newcommand\blfootnote[1]{%
  \begingroup
  \renewcommand\thefootnote{}\footnote{#1}%
  \addtocounter{footnote}{-1}%
  \endgroup
}

\onlineid{0}

\vgtccategory{Research}

\vgtcinsertpkg

\title{Practical Challenges of Progressive Data Science in Healthcare }

\author{Faisal Zaki Roshan \thanks{e-mail: faisalzaki@cmail.carleton.ca}\\ %
        \scriptsize Carleton University %
\and Abhishek Ahuja\thanks{e-mail: abhishekahuja@cmail.carleton.ca}\\ %
     \scriptsize Carleton University %
\and Fateme Rajabiyazdi\thanks{e-mail: fateme.rajabiyazdi@carleton.ca}\\ %
     \parbox{1.4in}{\scriptsize \centering Bruyere Research Institute \\ Carleton University}}

\abstract{
The healthcare system collects extensive data, encompassing patient administrative information, clinical measurements, and home-monitored health metrics. To support informed decision-making in patient care and treatment management, it is essential to review and analyze these diverse data sources. Data visualization is a promising solution to navigate healthcare datasets, uncover hidden patterns, and derive actionable insights. However, the process of creating interactive data visualization can be rather challenging due to the size and complexity of these datasets. Progressive data science offers a potential solution, enabling interaction with intermediate results during data exploration.
In this paper, we reflect on our experiences with three health data visualization projects employing a progressive data science approach. 

We explore the practical implications and challenges faced at various stages, including data selection, pre-processing, data mining, transformation, and interpretation and evaluation. 
We highlighted unique challenges and opportunities for three projects, including visualizing surgical outcomes, tracking patient bed transfers, and integrating patient-generated data visualizations into the healthcare setting.
 We identified the following challenges: inconsistent data collection practices, the complexity of adapting to varying data completeness levels, and the need to modify designs for real-world deployment. Our findings underscore the need for careful consideration of using a progressive data science approach when designing visualizations for healthcare settings.

} % end of abstract


\begin{document}


\firstsection{Introduction}

\maketitle

%% \section{Introduction} %for journal use above \firstsection{..} instead
Every day, a large amount of health data is collected and stored as a part of the healthcare system. These data include patient administrative information (i.e., demographic, visit records, appointments), patient data measured in the clinic (i.e., medical imaging, laboratory results, vital signs), and patient health data collected at home (i.e., blood pressure, heart rate, blood glucose). To optimize the healthcare industry's operational efficiency and enhance the delivery of healthcare services, clinicians, policymakers and management teams need to actively analyze healthcare data. However, this data is often large, messy, and difficult to interpret.

\blfootnote{IEEE VIS: Visualization \& Visual Analytics Conference, Progressive Data Analysis and Visualization Workshop 2024\\
St. Pete Beach, Florida, USA, October 13-18\\
Copyright held by authors.\\ }

Data visualization is a promising solution that can support analysts in uncovering unknown unknowns, i.e., questions that are only uncovered while visually browsing data, finding underlying patterns and relationships that otherwise would have remained hidden, and revealing patterns and discovering valuable insights that might have otherwise remained concealed~\cite{Bertini2010}.

Fortunately, the healthcare industry has had a growing interest in visually analyzing these datasets. However, despite the interest in visualizing these datasets, there are still challenges to visualizing and analyzing these datasets as the data is often large, incomplete, and collected in real-time. Progressive data science (PDS) could be a solution for data exploration in such large health data visualization systems to enable interaction with intermediate results~\cite{turkay2019progressivedatasciencepotential, fekete2019progressive}.

% inspired by the progressive DS pipeline introduced by [et al], we discuss the practical implications of different stages .

In this paper, we reflect on the opportunities and challenges we faced with our three visualization development projects in healthcare using the progressive data science approach.  We explore the practical implications and challenges faced at five stages of PDS, data selection, pre-processing, transformation, data mining, and interpretation and evaluation, as defined by Turkay et al.~\cite{turkay2019progressivedatasciencepotential}. Each project, from visualizing surgical outcomes to tracking patient bed transfers and integrating patient-generated data, highlights unique challenges such as inconsistent data collection practices, adapting to varying data completeness levels, and modifying designs for real-world deployment. Future work should focus on developing standardized methods and tools to streamline data collection, pre-processing, and interpretation and evaluation when using a progressive data science approach. 

% Taxonomy~\cite{Ulmer2023Survey}

\section{PROJECT\#1: SURGICAL OUTCOMES}

Many hospitals track and monitor various operational and patient health outcomes and clinical indicators to increase operational efficiency, aiming to enhance patient care, optimize resource utilization, minimize costs, and improve overall performance.  These indicators are essential for monitoring and evaluating hospital processes, identifying bottlenecks, and guiding evidence-based decision-making about enhancing the quality of care delivery. 

\subsection{Project Description and PDS}
In this project, we were invited to join forces with a local hospital to enhance to quality of surgical care. One pillar of this effort was to actively monitor and visually analyze the data. Thus, our goal was to design and develop an interactive data visualization system displaying surgical outcome measures 
% to improve patient surgical care and optimize operational efficiency by 
enabling healthcare providers and administrators to make informed decisions.

% \textbf{Dataset:} 
We were asked to design an interactive visualization displaying predefined Key Performance Indicators (KPIs) collected from pre-, per-, and post-surgical periods. 
We started designing low-fidelity and medium-fidelity data visualization prototypes using a sample dataset collected from two hospitals. Once we developed our designed interactive data visualization system, we deployed it into the healthcare setting. 

We used PDS to complete this project using PDS. We followed the five stages of data selection, pre-processing, transformation, data mining, and interpretation and evaluation.

\begin{figure*}[h]
\centering

\subfigure[Bar chart shows the distributions of 3 response options for\newline several surgery quality check indicators in the selected hospitals.]{\label{fig:Surgery-1}\includegraphics[width=85mm]{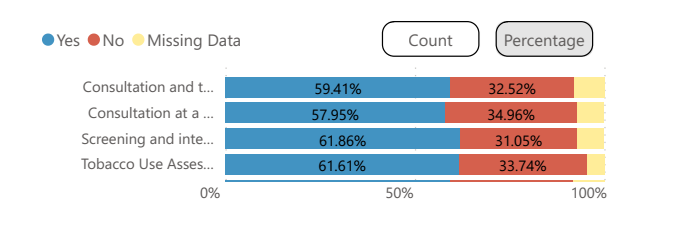}}
\subfigure[Bar chart shows the distributions of 5 response options for\newline several surgery quality check indicators in the selected hospitals.]{\label{fig:Surgery-2}\includegraphics[width=85mm]{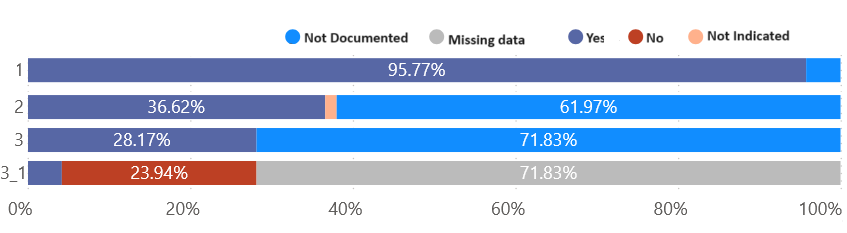}}
\caption{Changes to the response options for Project\#1.}
\label{fig:figures}
\end{figure*}

% port presents a comprehensive study on the development and implementation of a visualization tool designed to display key indicators in surgical outcomes.
% The primary objective of this study is

\subsection{Challenges}
% Progressive stages: 
% However, for this project, we were invited to the project while the data was being collected from a few centers. 
Below, we discuss the challenges we faced in this project using a progressive data science approach.

\textbf{C1: Data Selection:} We were invited to this project after parts of the data selection were done. This is a typical case in working with the healthcare industry, where the industry hires data scientists and data visualization experts to develop an interactive data visualization system for a dataset. These datasets are often collected in-house, in the hospital or by healthcare organizations.

However, when we started collaborating with hospitals for this project, only 2 hospitals out of over 50 hospitals started collecting data.
This introduced challenges in our process. 
Although we could work with parts of the data to start the design phase, it caused challenges in design later. 
Different centers collected data differently, which imposed design changes later in the process, consequently delaying the project.

% This caused delays in the project timeline.

\textbf{C2: Pre-processing:}
As mentioned earlier, when started working on this project, we only had access to a small dataset collected from two hospitals. As these centers were the ones who initiated the project, they were extra careful in collecting data, leaving a very limited number of missing data in the dataset. 
This was great at first to work with a wholesome dataset; however, we faced challenges when other centers started collecting data. These centers had many missing data items that needed to be dealt with in the design.

At first, the originating hospitals collected three response options (yes, no, missing data) to a given question (See Figure~\ref{fig:Surgery-1}).
When we gathered data from a few additional hospitals, we noticed other hospitals had two additional response options ``not documented'' and ``not applicable'', different from missing data.

This caused a challenge in both the design and development phases. 
We had a set colour scheme that was selected to be colour-blind safe and follow the institution's colour theme. However, to accommodate the two extra response options, we had to allocate two additional colours to represent all response options, which was challenging. The two additional colours needed to be easily distinguishable, colour-blind safe, and match the colour themes. This enforced some design changes in our visualization (See Figure~\ref{fig:Surgery-2}).

\textbf{C3: Transformation:}
Consequently, we had to modify the database to transform the newly formatted data to allow for the collection of two extra responses. This necessitated our team to re-run the previously tested components to ensure they work as intended.

% \begin{figure}
%     \centering
%     \includegraphics[width=1\linewidth]{figures/Pre-stage-1.png}
%     \caption{Bar chart shows the distributions of 3 response options for several surgery quality check indicators in the selected hospitals.}
%     \label{fig:Surgery-1}
% \end{figure}

% \begin{figure}
%     \centering
%     \includegraphics[width=0.9\linewidth]{figures/Pre-stage-3.png}
%     \caption{Bar chart shows the distributions of 5 response options for several surgery quality check indicators in the selected hospitals}
%     \label{fig:Surgery-2}
% \end{figure}

\textbf{C4: Data Mining:}
% The goal of the interpretation/evaluation step is to assess the performance of the data model calculations and the quality of data visualizations. 

In this project, we employed a progressive data science approach and displayed the results of the calculations and visualizations using parts of the data. These included examples such as calculating the rate of patients who experienced complications after surgery. 

% This started conversations with our medical collaborators on how to best display the results of the calculation.
% 2) Communication Challenges:

The early partial computations enabled conversations with our medical collaborators on how to best display the calculation results. 
Our team involved administrative staff, medical professionals, programmers, and the information technology team.
A challenge we faced was communicating a clear message that these results were not final.
Expectations were set on the visualizations to show the final results. The team members mentioned that the numbers displayed would not match the numbers they ran when calculating results. Thus, we had to spend a great deal of time and resources explaining the progressive process to those who were not familiar with this process.

\textbf{C5: Interpretation/Evaluation:}
Additionally, nearing the end of the project, since we used partial data and presented partial calculations, we realized some of the calculation formulas were not correctly used.
For example, to calculate the days spent by patients in the hospital, one can count day surgeries as 0 or not count them in the overall calculation (sum or average). This consideration changes the final results and the interpretation. Thus, we spent weeks revising such calculation formulas.

% Change in the pre-defined KPIs - some KPI definitions changed and some KPIs removed/added - this required changes in the backend as well as visualtions

% 1) Data Collection and Cleaning - Data Connection: DB to PBI
% Data variable selection
% Data updates on particular complication 
% Data collected from 1 hospital was displayed, 
% Challenges: Methods used to collect data entries differed for some hospitals. After other hospital data was added, new data items were introduced that affected the design
% Example: Some hospitals collected extra response options (Yes, No, Not applicable, Missing) – 
% Although progress was made in the design and development, one must investigate if the extra time and resources to fix the additional….

\section{PROJECT\#2: PATIENT BED TRANSFER STATES}

A local hospital contacted us to find a solution for tracking and analyzing their patient smart bed transfers in the hospital to monitor and enhance the transfers.
% for the hospital administrators. 
% These log files included data transmitted over the CAN (Controller Area ...
        
\begin{figure*}[t]
\centering

\subfigure[Hospital bed transfer visualization dashboard while the calculations\newline in progress.]{\label{fig:Alta-Before}\includegraphics[width=85mm]{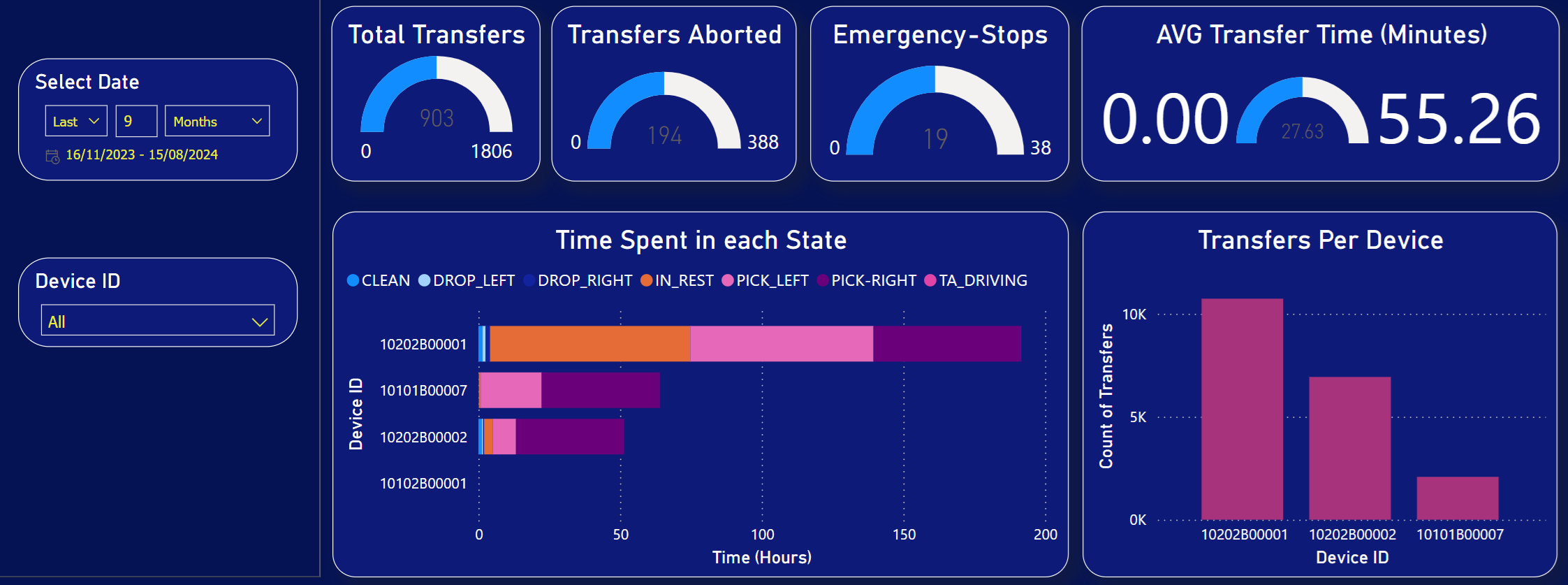}}
\subfigure[Hospital bed transfer visualization dashboard after the calculations\newline are finalized.]{\label{fig:Alta}\includegraphics[width=85mm]{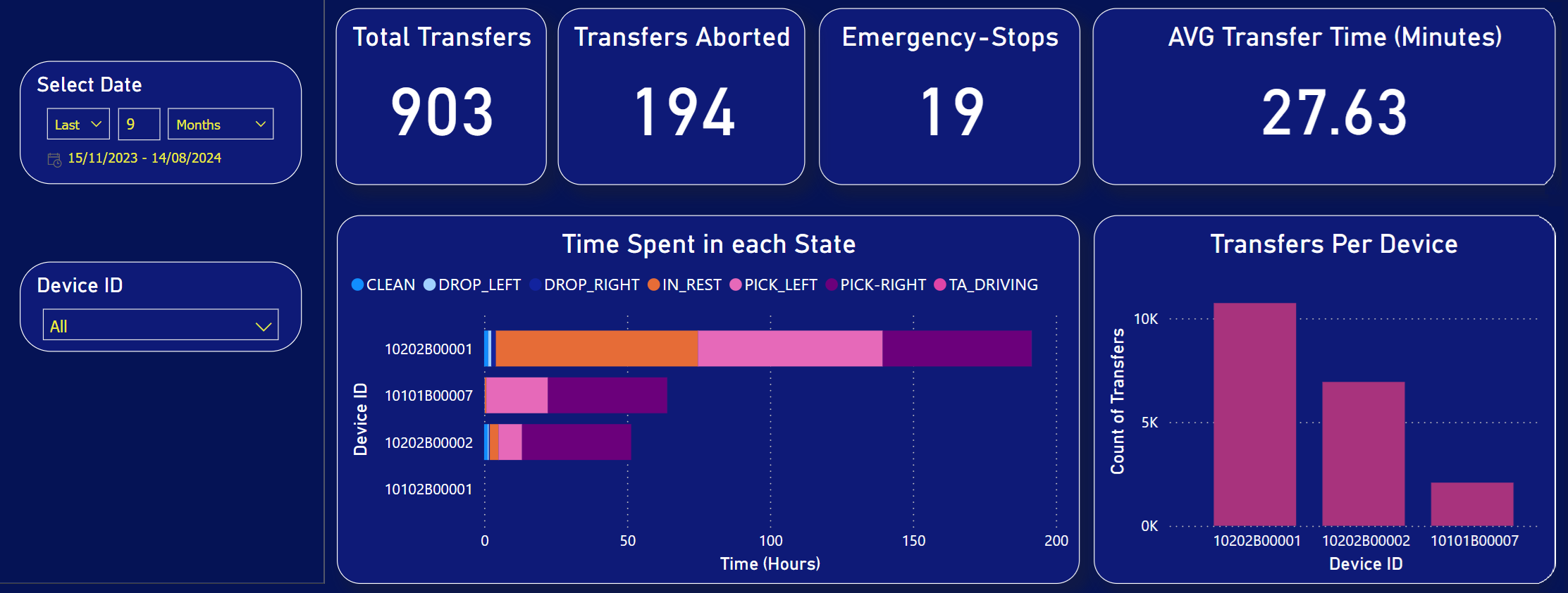}}
\caption{Progressive demonstrations of the bed transfer reports for Project\#2.}
\label{fig:Alta-PR2}
\end{figure*}

\subsection{Project Description and PDS}
Our goal was to design and develop an interactive data visualization system of the beds' sensor log files generated by smart bed transfers to enhance delivery and operational efficiency. 

\subsection{Challenges}
Below, we discuss the challenges we faced in this project using a
progressive data science approach.

\textbf{C1: Data Selection:}
A small sample retrospective dataset was shared with our team to start our design steps. 
The dataset was a collection of bed sensor logs when transferring a patient between units in a hospital. 
Using this small dataset, we sketched and designed several visualization representations. Upon sharing the designs with the healthcare provider team, we received feedback on the design and modified the design in an iterative process.
% We have not yet had the opportunity to work or test our visualizations with the real-world prospective data on this project, we anticipate changes to be requested.
Upon extensive discussions with healthcare providers and stakeholders, we selected key visualizations that best represent smart bed usage data for integration into the healthcare management system.

\textbf{C2: Pre-processing:} To transform our sample raw data into a structured format, we used a Python script and parsed the log files, extracting pivotal information while discarding irrelevant details. 
We then used the cleaned data as input to create a Microsoft Power BI data model.

\textbf{C3: Transformation:}
In Power BI, we further transformed the data to enrich the data model and facilitate the calculation of KPIs critical to the project's objectives.
 When we shared these KPIs with our collaborators, we received feedback that the indicators did not display the information they anticipated. We then had to make changes to the calculation of KPIs.
 
\textbf{C4: Data Mining:}
% As mentioned, we started our project with a small dataset of sample data. Some parts of this dataset had missing data input or data variables that were not collected data
We started the project with a small dataset and designed the data visualizations accordingly. However, data at a large city hospital scale will exponentially grow, particularly in the case of bed transfer that happens multiple times a day for many patients.

Over the course of the project, more data was accumulated and displaying the final results was no longer possible. The hospital data recording systems are often old and slow. 
Thus, upon getting feedback from the hospital staff and administrators, we adopted our design to a progressive data representation. We will display a partial calculation of the KPIs (See Figure~\ref{fig:Alta-Before}) using a Gauge chart (i.e., a speedometer or dial chart) to show partial data and update the charts as data are collected. 
Once the numbers are finalized, we will display the final numbers (See Figure~\ref{fig:Alta}). We are aware that showing a partial number may be risky as it may be interpreted wrongfully. This needs to be clearly stated in the dashaboards.

% (. As an alternative design, we are planning to use a Gauge chart (i.e., a speedometer or dial chart) to show partial data and update the charts as data are collected.
% \begin{figure}[h]
%     \centering
%     \includegraphics[width=\linewidth]{figures/Alta-2.png}
%     \caption{}
%     \label{fig:Alta-Before}
% \end{figure}
% \begin{figure}[h]
%     \centering
%     \includegraphics[width=\linewidth]{figures/Alta.png}
%     \caption{Hospital bed transfer visualization dashboard after the calculations are finalized.}
%     \label{fig:Alta}
% \end{figure}

\textbf{C5: Interpretation/Evaluation:} 
% Design and Deployment Adjustments for Smart Bed Insights Dashboard
% The selected visualizations included:
% Patient Transfer Metrics Visualization: Designed to show real-time data on patient transfers.
%  Bed Occupancy Rates Visualization: Focused on displaying current and historical bed occupancy.
%  Device Operational Statuses Visualization: Provided insights into the operational status and efficiency of smart bed

 % Interpretation/Evaluation: Design and Deployment Adjustments…
% The challenge here was the
 This project started out with a loosely defined set of requirements and interpretation of data. The visualizations needed to be modified even at the final stages of the project, this caused extensive re-work on the calculations, which caused more changes in the backhand.
 
\section{PROJECT\#3: PATIENT-GENERATED DATA}

Collecting patient-generated data is becoming increasingly common in chronic disease management~\cite{fox2013tracking}. 
Patients use various tracking tools to collect health and lifestyle data~\cite{ancker2015invisible}. 
However, most of these tracking tools are not designed to fully meet patients' and healthcare providers' expectations~\cite{Choe2014}
and do not support reviewing patient-generated data with healthcare providers during clinical visits. 
One way to support patients in discussing their data with healthcare providers is effective visualization of patient-generated data collections.

\subsection{Project Description and PDS}
We were approached by a group of healthcare providers from a local hospital who are involved in the care of chronic patients to explore if, and how, to design technology that can enhance the presenting and reviewing of patient-generated data during a clinical visit. To answer this question, we took an iterative approach with the involvement of patients and healthcare providers and designed various visualizations representing the patient-generated data collections~\cite{RajabiyazdiWicked2021,rajabiyazdiGI2020}.

\subsection{Challenges}

Below, we discuss the challenges we faced in this project using a
progressive data science approach.

\textbf{C1: Data Selection:}
Both healthcare providers and patients agree that patient-generated data, could be used by patients to make 
informed decisions to improve their quality of life, and could aid providers in making decisions about patient ongoing care~\cite{zhu2016sharing}. 
There are existing technologies for tracking and 
visualizing health data such as sleep (e.g.,~\cite{Choe2015Sleep}), physical activity (e.g.,~\cite{Consolvo2008}), and blood sugar level (e.g., ~\cite{Brzan2016}). We asked eight patients to bring a sample of their data and share it with us.

\textbf{C2: Pre-processing:} 
Some patients recorded and maintained
their data on paper mixed with other personal information. 
Thus, as researchers, we needed to clean and transform hand-collected data into digital forms. Some of the patients who used
apps or tools to collect data did not have easy
access to their data as some of these tools did not
provide an easy way to export or share their data.
Thus, for some of the patients, we only had access to partial data.

\textbf{C3: Transformation:}
Data collected by patients came in different formats, chosen personally by each patient.
To include the data collected from patients in the electronic medical record, a certain format needs to be followed.
We needed to reformat the data input to acceptable forms that are readable by medical record systems.  

\textbf{C4: Data Mining:}
We designed various individually tailored visualizations based
on patient health data they collected at home.
Upon extensive discussions with the healthcare providers, as a group we selected four visualizations that best represent patient data for incorporation into the provincial care plan platform.

One design was dedicated to showing blood pressure data, one design to representing stress level data, and two designs to displaying blood glucose data (one for patients who use an insulin pump and one targeting diabetes patients who use an insulin pump). 

\begin{figure*}[t]
\centering

\subfigure[Glucose data visualizations in progress.]{\label{fig:Glucose-B}\includegraphics[width=78mm]{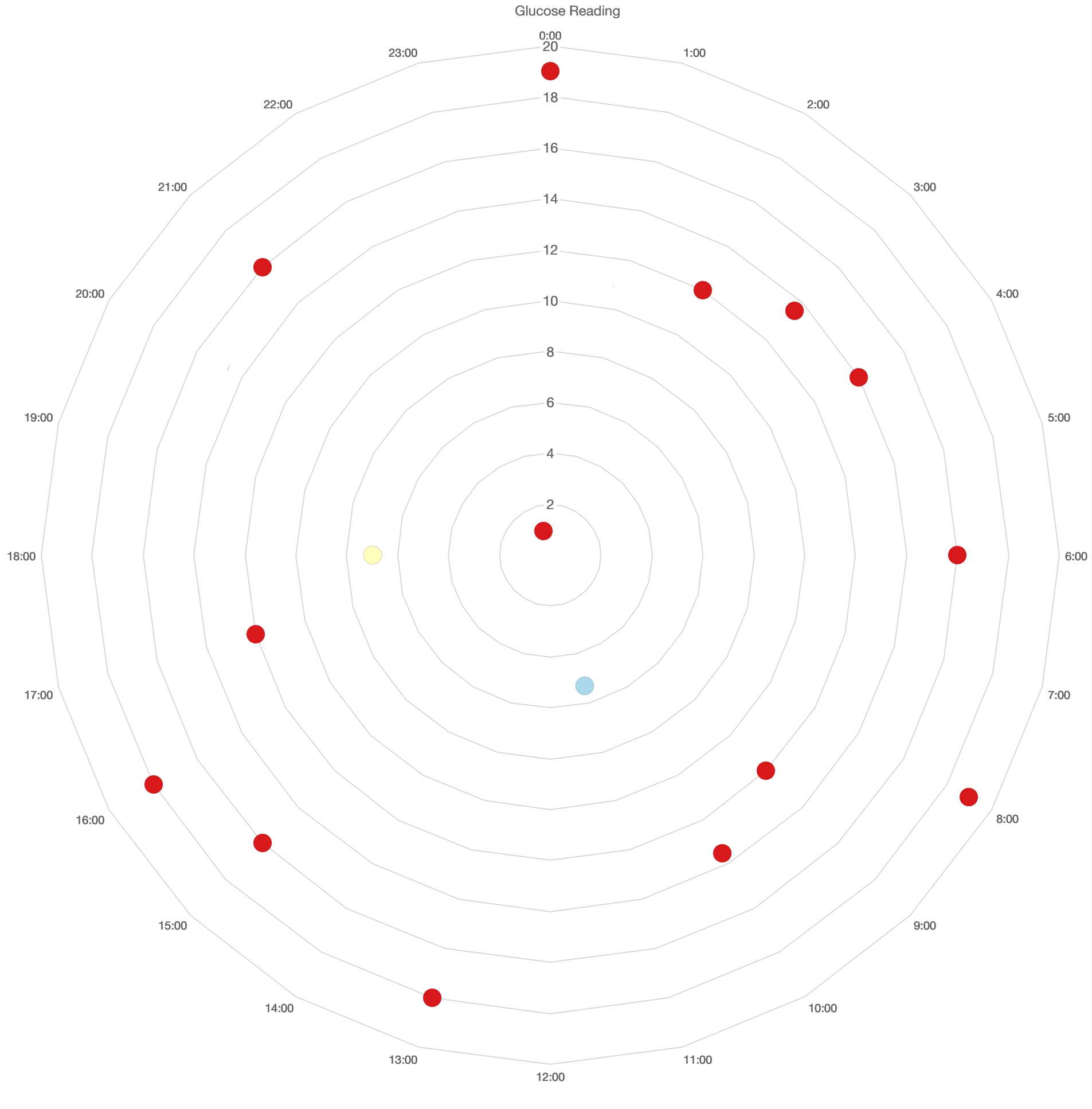}}
\subfigure[Glucose data visualizations finalized.]{\label{fig:Glucose-A}\includegraphics[width=80mm]{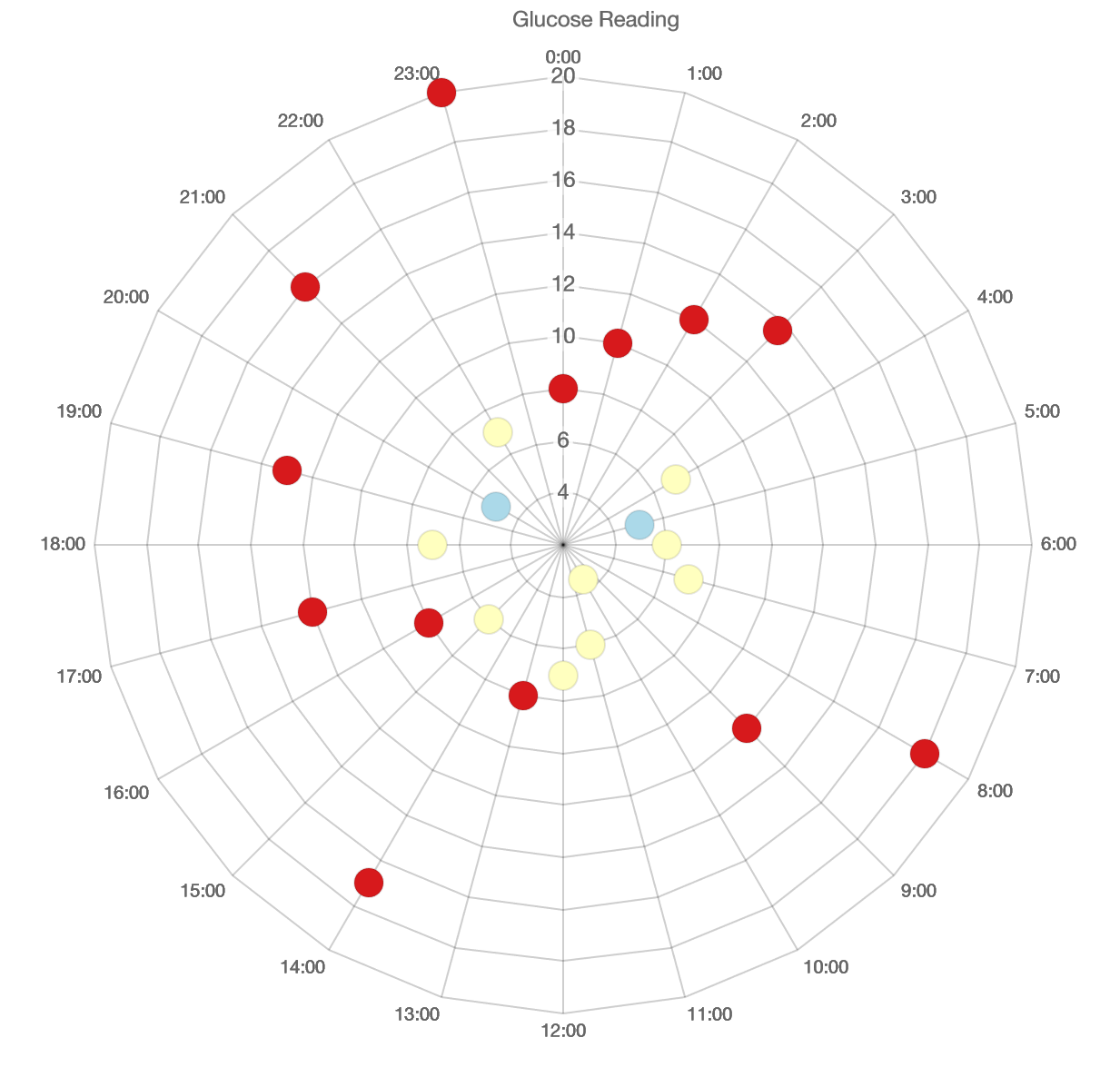}}
\caption{Glucose data visualizations progression for Project\#3.}
\label{fig:Glucose}
\end{figure*}

One design displays blood glucose from a Continuous Glucose Monitor for patients using an insulin pump machine. The continuous measures of glucose can be accumulated, making a large dataset when collected over months. 
Displaying this large dataset visually in a graph can take a large computational effort. Each data point needs to be placed on the graph in the correct placement and take an associated colour code demonstration (red: too high or too low glucose; yellow: borderline; and blue: normal glucose).
From our discussions with healthcare providers, we gathered they sometimes do not need to review all this data in detail particularly during a short clinical visit. Thus, one proposed solution is to use PDS and demonstrate parts of the data (See Figure~\ref{fig:Glucose-B}) before finalizing all data placement on the chart (See Figure~\ref{fig:Glucose-A}). If healthcare providers have a concern and need to further examine the data, they may wait for the full data to be processed. 

% \begin{figure}
%     \centering
%     \includegraphics[width=0.8\linewidth]{figures/GlucoseRadar-after-1.png}
%     \caption{Glucose data visualizations in progress.}
%     \label{fig:Glucose-B}
% \end{figure}

% \begin{figure}
%     \centering
%     \includegraphics[width=0.8\linewidth]{figures/GlucoseRadar-before.png}
%     \caption{Glucose data visualizations finalized.}
%     \label{fig:Glucose-A}
% \end{figure}

\textbf{C5: Interpretation/Evaluation:} 
% \textbf{Data Mining \& Interpretation/Evaluation:} 
% Color change when deployment - minimal design and coding
% \textbf{Data Mining } 
Once we confirmed the design, the next step was to deploy them. However, we had to make changes to our original designs in preparation for developing and integrating the designs into the provincial care plan. Some of the libraries we used for rendering the visualizations were not in the accepted list of libraries for deploying into the healthcare system.
This challenge was not raised when we showed the design and development in a progressive manner to the stakeholders.
At this stage, we had to change some of the designs to accommodate the available visualization and development libraries, which caused further delays in the project.

% Such integration comes with many challenges,
% including understanding technical constraints and communicating data mappings     
% when handing-off visualization designs for development~\cite{Walny2020}.
% Thus, visualization designers need to be adaptable to changes, particularly when these designs need to be compatible with a given healthcare system.
% Deployment issue - libraries not available 

\section{CHALLENGES and OPPORTUNITIES}
Here we discuss the practical challenges and opportunities we identified when employing progressiveness in designing and developing data visualization dashboards in the healthcare setting. Although taking on a progressive approach was instrumental in the rapid progress of the visualization projects, it introduced some challenges. Particularly in the Data collecting and cleaning, Data Mining, and Interpretation/Evaluation phases. Often these challenges introduced issues in rendering and deploying the data visualizations. To solve these challenges, we needed to make changes to the design of the visualization, a step backward from development or deployment to the design process.

 % Designing and re-designing

 % Communication and setting the right expectations

 % Deployment challenges + \cite{Zoumpatianos2018}

\section{Conclusion}
Our exploration of progressive data science in healthcare reveals several challenges and potentials. By enabling real-time interaction with intermediate results, progressive data science can speed up the progression of such projects. However, this approach also introduces unique challenges, particularly in data selection, pre-processing, and consistent communication with stakeholders.

Each project we undertook—visualizing surgical outcomes, tracking patient bed transfers, and integrating patient-generated data—highlighted the need for flexibility and iterative design. Inconsistent data collection practices, varying completeness levels, and adapting designs for deployment in real-world settings necessitated continuous adjustments and stakeholder engagement.

Future work should focus on developing standardized methods and tools to streamline data collection, pre-processing, and visualization when using a progressive data science approach. Additionally, fostering strong collaboration and clear communication among all stakeholders is crucial for successful implementation. By addressing these challenges, we can harness the potential of progressive data science to enhance healthcare delivery and outcomes.

\acknowledgments{
The authors wish to thank all patient collaborators and the healthcare providers who provided input in these projects.
}

\balance
\bibliographystyle{abbrv-doi}

\bibliography{template}
\end{document}